# Origin of the hump anomalies in the Hall resistance loops of ultrathin SrRuO$_3$/SrIrO$_3$ multilayers

Lin Yang,[1] Lena Wysocki,[1] Jörg Schöpf,[1] Lei Jin,[2] András Kovács,[2] Felix Gunkel,[3]
Regina Dittmann,[3] Paul H. M. van Loosdrecht,[1] and Ionela Lindfors-Vrejoiu[1]

[1]*University of Cologne, Institute of Physics II, 50937 Cologne, Germany*

[2]*Ernst Ruska-Centre for Microscopy and Spectroscopy with Electrons,
Forschungszentrum Jülich GmbH, 52425 Jülich, Germany*

[3]*PGI-7, Forschungszentrum Jülich GmbH, 52428 Jülich, Germany*

(Dated: September 22, 2020)

The proposal that very small Néel skyrmions can form in SrRuO$_3$/SrIrO$_3$ epitaxial bilayers and that the electric field-effect can be used to manipulate these skyrmions in gated devices strongly stimulated the recent research of SrRuO$_3$ heterostructures. A strong interfacial Dzyaloshinskii-Moriya interaction, combined with the breaking of inversion symmetry, was considered as the driving force for the formation of skyrmions in SrRuO$_3$/SrIrO$_3$ bilayers. Here, we investigated nominally symmetric heterostructures in which an ultrathin ferromagnetic SrRuO$_3$ layer is sandwiched between large spin-orbit coupling SrIrO$_3$ layers, for which the conditions are not favorable for the emergence of a net interfacial Dzyaloshinskii-Moriya interaction. Previously the formation of skyrmions in the asymmetric SrRuO$_3$/SrIrO$_3$ bilayers was inferred from anomalous Hall resistance loops showing humplike features that resembled topological Hall effect contributions. Symmetric SrIrO$_3$/SrRuO$_3$/SrIrO$_3$ trilayers do not show hump anomalies in the Hall loops. However, the anomalous Hall resistance loops of symmetric multilayers, in which the trilayer is stacked several times, do exhibit the humplike structures, similar to the asymmetric SrRuO$_3$/SrIrO$_3$ bilayers. The origin of the Hall effect loop anomalies likely resides in unavoidable differences in the electronic and magnetic properties of the individual SrRuO$_3$ layers rather than in the formation of skyrmions.

## I. INTRODUCTION

Topologically protected magnetic whirls, dubbed as magnetic skyrmions, are considered to be ideal candidates for the potential application in future data storage [1]. This primarily derives from their small size and room temperature stability [2–4], low energy consumption [3–7], and topological protection [8–10]. Epitaxial perovskite oxide heterostructures, such as SrRuO$_3$/SrIrO$_3$ are considered to have strong interfacial Dzyaloshinskii–Moriya interaction (DMI) due to the broken spatial inversion symmetry and the strong spin-orbit coupling in SrIrO$_3$ and it was reported that Néel skyrmions form in these heterostructures [11, 12]. The insulating nature of perovskite oxide heterostructures, such as ultrathin SrRuO$_3$/SrIrO$_3$, makes them promising systems in terms of electric field manipulation as well as the ability to engineer their magnetic properties. Magnetic skyrmions can in principle be observed in real space by magnetic force microscopy (MFM) [13], Lorentz transmission electron microscopy (LTEM) [14], scanning transmission x-ray microscopy (STXM) [3], spin-polarized scanning tunneling microscopy (SP-STM) [15], x-ray magnetic circular dichroism based photoemission electron microscopy (XMCD-PEEM) [16], spin-polarized low energy electron microscopy (SPLEEM) [17], and in reciprocal-space by small-angle neutron scattering (SANS) [18] and resonant small-angle X-ray scattering (SAXS) [19]. However, for epitaxial oxide heterostructures, by using these techniques, the direct observation of sub-100 nm size skyrmions and their characterization become extremely challenging. Therefore, the possibility of examining the formation of skyrmions by magnetotransport measurements is very attractive as Hall resistivity investigations are rather easy to perform in any solid state research laboratory. Recently, there were many reports in which the formation of skyrmions was inferred from the observation of humplike anomalies of Hall resistance loops that were attributed to the manifestation of the topological Hall effect (THE). This is the case of quite a few reports related to epitaxial SrRuO$_3$ heterostructures and to bare SrRuO$_3$ thin films [20–22]. Matsuno *et al*. [11] attributed the observation of such features of Hall loops measured for ultrathin ferromagnetic SrRuO$_3$ (thinner than 6 pseudocubic unit cells (uc)) capped by 2 uc SrIrO$_3$ thick layer to the formation of skyrmions. Many similar publications followed shortly. There were reports of the humplike features observed in Hall resistance loops of a variety of SrRuO$_3$ based samples: SrRuO$_3$/SrIrO$_3$ multilayers with relatively thick layers (10 uc thick SrRuO$_3$) [23], BaTiO$_3$/SrRuO$_3$ bilayers [24], SrRuO$_3$ (5 uc)/ SrIrO$_3$ (2 uc) in which the iridate layer was the bottom layer on the SrTiO$_3$ substrate [25], SrRuO$_3$ (8 uc)/BaTiO$_3$ (2 uc) bilayers on SrTiO$_3$ [26], SrIrO$_3$ (2 uc)/SrRuO$_3$ (10 uc) bilayers for which MFM experiments were also performed [12], or relatively thick SrRuO$_3$ (3-6 nm) films grown in low oxygen pressure [27]. Different mechanisms for the occurrence of the interfacial DMI were proposed in these papers, adapted to the particular interfaces and sample peculiarities.

However, the interpretation of the observed humps in anomalous Hall effect (AHE) resistance loops as a fingerprint of the THE contribution due to skyrmions is currently under debate. Other reports addressed the possi-ble role played by SrRuO$_3$ layer inhomogeneity, such as, Ru/O vacancies [28], thickness variations [29–31], crystal structure distortions [26, 32], and intermixing [33] in the occurrence of the THE-like features of the AHE loops. This division of opinions concerning the origin of the THE-like structures of the Hall resistance loops calls for a careful analysis and understanding of the electronic and magnetotransport properties of SrRuO$_3$-based heterostructures. We stress that there are no direct measurements of the magnitude of the interfacial DMI in such epitaxial SrRuO$_3$-based heterostructures, but only the theoretical proposal from Ref. [11], which does not however provide any quantitative microscopic description of how the DMI is generated at the SrIrO$_3$/SrRuO$_3$ interfaces. There exists very little insight in the interfacial DMI at epitaxial oxide interfaces [34], although hints for the existence of an interfacial DMI in SrIrO$_3$ (2 uc)/SrRuO$_3$ (10 uc) bilayers were inferred from the analyses of the magnetic domain wall chirality [12]. We previously studied asymmetric SrZrO$_3$/SrRuO$_3$/SrIrO$_3$ and SrHfO$_3$/SrRuO$_3$/SrZrO$_3$ multilayers in which we aimed to observe the possible effects of the net interfacial DMI on the magnetotransport properties and magnetic domain formation [35]. However, these SrRuO$_3$ multilayers, with insulating spacers, were magnetically only very weakly coupled [36] and did not permit a conclusive investigation of the magnetic domains by magnetic force microscopy.

Here we deliberately considered SrIrO$_3$/SrRuO$_3$/SrIrO$_3$ epitaxial trilayers and multilayers with several repeats of the trilayer, in order to have interfaces as symmetric as possible in this material system. We aimed to eliminate, or at least minimize, the role of interfacial DMI. In a perfectly symmetric ultrathin film heterostructure the interfacial DMI should cancel out. However, for epitaxial interfaces of perovskite oxides (ABO$_3$), the interfaces are likely to be asymmetric due to the AO/BO$_2$ stacking imposed by epitaxial growth, due to asymmetric intermixing or different oxygen octahedral rotations (OOR) angles at the upper and lower interface. For example, the strong influence of the type of interface stacking on the physical properties of perovskite oxide heterostructures was recently demonstrated for SrIrO$_3$-La$_{0.7}$Sr$_{0.3}$MnO$_3$ bilayers [37]. For our trilayers and multilayers, because A = Sr for both SrRuO$_3$ and SrIrO$_3$, the interfaces are either of the type SrO/IrO$_2$//SrO/RuO$_2$ or IrO$_2$/SrO//RuO$_2$/SrO, and from this viewpoint the interfaces are equivalent.

Prior investigations demonstrated that SrRuO$_3$ layers separated by 2 uc thick SrIrO$_3$ non-magnetic layers are magnetically decoupled [38]. Therefore, the overall conditions in these SrIrO$_3$/SrRuO$_3$/SrIrO$_3$ multilayers strongly disfavor the formation of Néel skyrmions. The trilayer SrIrO$_3$/SrRuO$_3$/SrIrO$_3$ samples did not exhibit any humplike anomalies in the Hall effect loops. In contrast, humplike anomalies were observed over a large temperature range in Hall effect loops of nominally symmetric multilayer samples, in which a SrRuO$_3$/SrIrO$_3$ bilayer was stacked 3 or 6 times. The Hall effect loops with hump anomalies can be the result of inhomogeneous magnetic and electronic properties of the SrRuO$_3$ layers in the multilayers. The inhomogeneous properties possibly arise from layer thickness variation, different degree of intermixing of Ir on the Ru-site, and oxygen octahedron deformations that can be different for the SrRuO$_3$ layers next to the substrate and for the layers at the top of the multilayer [39].

## II. METHOD

### A. Sample growth

The heterostructures studied here, SrIrO$_3$/[SrRuO$_3$/SrIrO$_3$]$_m$ ($m$ = 1, 6) were grown on SrTiO$_3$(001) by pulsed-laser deposition (PLD) using a KrF excimer laser ($\lambda$ = 248 nm). SrTiO$_3$(001) single-crystal substrate was used for the deposition after NH$_4$F-buffered HF etching for 2 - 2.5 min and annealing at 1000 °C for 2 hours in air. The oxygen partial pressure and deposition temperature were optimized at 0.133 mbar and 650 °C for the growth of all the layers. The pulse repetition rate of the laser was 5 Hz and 2 Hz for the SrRuO$_3$ layers and SrIrO$_3$ layers, respectively. The growth of each layer was monitored by high oxygen pressure reflective high-energy electron diffraction (RHEED). The thickness of each SrRuO$_3$ layer is nominally 6 uc and the thickness of each SrIrO$_3$ layer is nominally 2 uc (1 uc layer is $\sim$ 0.4 nm thick). The samples were cooled in 100-200 mbar oxygen atmosphere from 650 °C down to room temperature with a rate of 10 °C/minute. A multilayer with $m$ = 3, SrIrO$_3$/[SrRuO$_3$/SrIrO$_3$]$_3$, was grown in a second RHEED-PLD system (made by SURFACE systems+ technology GmbH und Co. KG), under similar growth conditions, except for a higher laser fluence and target-to-substrate distance. The properties of the sample with $m$ = 3, along with the properties of a second trilayer ($m$= 1) reference sample made in this PLD system, are discussed in the supplementary online material (see section 2 and section 3).

### B. Sample characterization

The surface morphology of our samples was characterized by atomic force microscopy (AFM), as shown in the supplementary material. The microstructure of the multilayers, in terms of sharpness of the interfaces, layer thickness and element distribution, was investigated by high-angle annular dark field scanning transmission electron microscopy (HAADF-STEM) of cross section specimens. The distribution of the atomic elements was observed with high resolution energy dispersive X-ray spectroscopy (EDX). Both STEM and EDX were performed using an electron probe aberration corrected FEI Titan 80-200 ChemiSTEM microscope equipped with in-column EDX detectors. Hall effect measurements were carried out in the four-point geometry (van der Pauw), with permutating contacts for antisymmetrization. Hall resistance loops were recorded both with a Physical Property Measurement System (PPMS, Quantum Design Inc.) and with a home-made setup. The home setup enables the simultaneous measurement of transverse Hall resistance and magneto-optical Kerr effect (MOKE). The polar MOKE studies were performed with the magnetic field applied perpendicular to the thin film surface with incoherent light from a Xe lamp. The probe wavelength was chosen individually for each sample to reduce the contributions of optical artifacts, such as interference effects that can be present in heterostructures with ultrathin films of dissimilar oxides. Light of 491-520 nm wavelength was used for the $SrIrO_3/[SrRuO_3/SrIrO_3]_1$ trilayers, 630 nm wavelength was used $SrIrO_3/[SrRuO_3/SrIrO_3]_3$, and 610 nm wavelength was used $SrIrO_3/[SrRuO_3/SrIrO_3]_6$ multilayer.

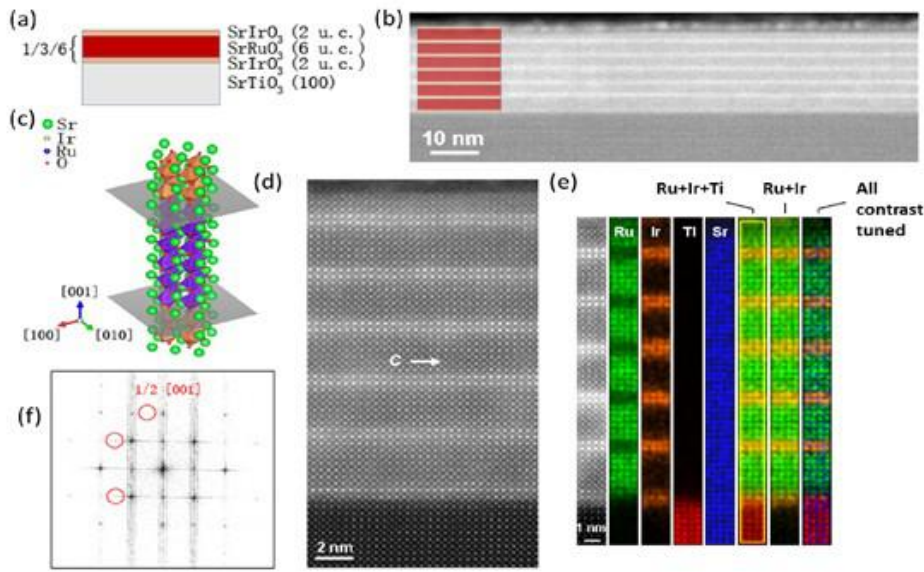

FIG. 1. Microstructure investigations by scanning transmission electron microscopy. (a) Schematics of sample $SrIrO_3/[SrRuO_3/SrIrO_3]_m$ ($m$ = 1, 3, 6) grown on $SrTiO_3$(100) substrates. (b) An overview HAADF-STEM micrograph of sample $SrIrO_3/[SrRuO_3/SrIrO_3]_6$ indicates the layers are uniform (except for the top layer that was damaged during the FIB processing of the specimen). (c) Schematics of the structure of the trilayer $SrIrO_3/[SrRuO_3/SrIrO_3]_1$, for which a 6 uc $SrRuO_3$ layer is inserted between two $SrIrO_3$ layers (both 2 uc thick). Green, orange, blue, and red dots represent Sr, Ir, Ru, and O atomic column positions, respectively, in the crystal structure drawn using VESTA [40]. In (d) and (e) high magnification micrographs show the quality of the interfaces. EDX elemental mapping across the entire stacks shown in (e) probed the uniformity of chemical element distribution. (f) FFT pattern obtained from the image shown in (d), which shows the spots due to the reflections originating from the orthorhombic distortion (marked by red circles), and confirms the in-plane $c$-axis orientation of the layers [white arrow in (d)].

The magnetic moment of the samples was measured as a function of temperature and magnetic field using a superconducting quantum interference device (SQUID) magnetometer (MPMS XL-7 from Quantum Design). The magnetic background due to the diamagnetic $SrTiO_3$ substrates was subtracted from the total magnetic response and often also corrections for a ferromagnetic impurity contribution had to be applied [35].

### III. RESULTS AND DISCUSSION

#### A. Microstructure investigations

The results of microstructure investigations by HAADF-STEM and high-resolution EDX are summarized in **Fig.** 1 The overall structure of the heterostructures under study here, $SrIrO_3/[SrRuO_3/SrIrO_3]_m$ ($m$ = 1, 3, 6) is shown in **Fig.** 1(a), whereas the symmetric unit cell structure of the repetitive trilayer building block is depicted in **Fig.** 1(c). **Fig.** 1 (b) and **Fig.** 1 (d) show cross-sectional STEM images of the microstructure of the $SrIrO_3/[SrRuO_3/SrIrO_3]_6$ multilayer at low and high magnification, respectively. The stacking starts with a $SrIrO_3$ and individual $SrRuO_3$ and $SrIrO_3$ layers have thicknesses that match fairly well the expected thickness values from the *in situ* RHEED monitoring of the layer deposition (see supplementary material, Fig. S1(c)). The high resolution image [**Fig.** 1(d)] indicates coherent epitaxial growth of the layers, as no dislocations were detected across the entire stack in the investigated TEM specimen. The high resolution EDX mapping of the elements across the entire stacks [**Fig.** 1(e)] indicates that Ru and Ir are distributed as expected from the designed growth sequence starting with a SrIrO$_3$ layer as the first layer on the substrate. We could not analyze quantitatively the exact stoichiometry of the individual layers. Line profiles confirmed that the SrRuO$_3$ layers are about 6-7 uc thick (as the number of individual Ru-O$_2$ planes varies between 6 and 7), while the SrIrO$_3$ layers are 2-3 uc thick (as the number of individual Ir-O$_2$ planes varies between 2 and 3) (see supplementary material, Fig. S2). Concerning the intermixing, because the individual SrIrO$_3$ layers are much too thin, no quantitative analyses of the possible intermixing at the interfaces with the SrRuO$_3$ layers are feasible. Achieving atomic resolution in EDX investigations, due to the electron beam channeling, volume and spectrum background effects, is very problematic.

The structure was analyzed by fast Fourier transform (FFT) images [**Fig.** 1(f)] which confirm the in-plane $c$-axis orientation of the layers (see white arrow) and demonstrate the expected orthorhombic distortions (due to A-site atom displacements of the pseudocubic perovskite ABO$_3$) by the presence of extra reflections, marked by the red circles in **Fig.** 1(f).

### B. Magnetic properties

We measured the dependence of the out-of-plane total magnetic moment as a function of temperature, under zero-field cooling (ZFC, measured while heating up in 0.1 Tesla (T) after cooling the sample with no applied field) and field cooled (FC) with a 0.1 T field applied perpendicular to the sample surface. The results for the SrIrO$_3$/[SrRuO$_3$/SrIrO$_3$]$_1$ trilayer and the SrIrO$_3$/[SrRuO$_3$/SrIrO$_3$]$_6$ multilayer are summarized in **Fig.** 2. For the trilayer sample, the Curie temperature ($T_c$) is 126 K, which was determined from the derivative of the FC magnetic moment curve as a function of temperature (see inset in **Fig.** 2(a)). The $T_c$ of the 6 uc thick SrRuO$_3$ layer of this sample is lower than for the bulk SrRuO$_3$ single crystals ($T_c$ = 160 K), which is typical for ultrathin films, due to epitaxial strain and disorder and stoichiometry effects, which are more pronounced the thinner the SrRuO$_3$ layers are[41]. The magnitude of the magnetic moment for SrIrO$_3$/[SrRuO$_3$/SrIrO$_3$]$_6$ is almost 6 times as large as SrIrO$_3$/[SrRuO$_3$/SrIrO$_3$]$_1$ (see the red dotted curve in **Fig.** 2(b)), corresponding to the magnetic volume relation of these two samples. Apparently two transitions at temperatures $T_{c1}$ (120 K) and $T_{c2}$ (140 K) occur for the SrIrO$_3$/[SrRuO$_3$/SrIrO$_3$]$_6$ epitaxial multilayers. We assume that the occurrence of two transition temperatures originates from the inhomogeneous magnetic properties of the SrRuO$_3$ layers. Most likely the six SrRuO$_3$ layers of the SrIrO$_3$/[SrRuO$_3$/SrIrO$_3$]$_6$ have all slightly different Curie temperatures distributed in the interval between $T_{c1}$ and $T_{c2}$. Comparing with the transition temperature of the trilayer sample, which is 126 K, we are led to consider that the bottom most SrRuO$_3$ layer has the lowest Curie temperature, while the top SrRuO$_3$ layers have the largest Curie temperature. It is likely that the bottommost SrRuO$_3$ layers are most affected by the epitaxial strain and oxygen octahedral accommodation to the conditions of the SrTiO$_3$ substrate, resulting in suppressed Curie temperature. The topmost SrRuO$_3$ layers of the multilayer may have structures that are more relaxed towards the bulk SrRuO$_3$ structure, approaching the OOR values of the bulk, and thus have larger Curie temperature.

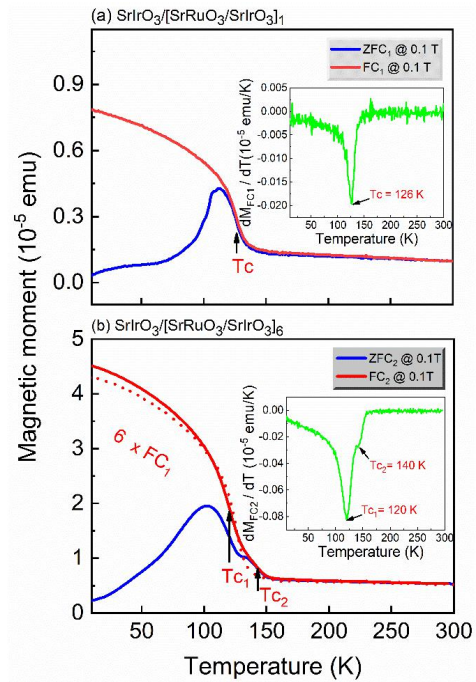

FIG. 2. Temperature dependence of the magnetic moment of the samples (a) SrIrO$_3$/[SrRuO$_3$/SrIrO$_3$]$_1$ and (b) SrIrO$_3$/[SrRuO$_3$/SrIrO$_3$]$_6$ under zero field cooling (ZFC, blue plot) and field cooling (FC, red plot, 0.1 T applied perpendicular to the sample surface) conditions. The dotted red curve shown in (b) is the FC$_1$ curve of the trilayer sample from (a) multiplied by 6 and plotted for the sake of comparison. The insets in (a) and (b) show the first derivative of the magnetization with respect to temperature, used to determine the Curie transition temperatures of the SrRuO$_3$ layers.

Two ferromagnetic transition temperatures were reported recently for (SrRuO$_3$)$_n$/(SrIrO$_3$)$_n$ superlattices with ultrathin individual layers ($n \leq 3$) [42]. The high temperature transition, occurring also at 140 K as for our samples, was attributed to the interesting possibility that the ultrathin SrIrO$_3$ layers undergo a canting antiferromagnetic transition. This transition vanished for the superlattices with thicker layers, $n \geq 4$. As stressed in this reference, no X-ray circular magnetic dichroism spectroscopy (XCMD) measurements at the Ru and Ir edges have been performed yet to test this proposal. There are however XMCD studies of LaMnO$_3$/SrIrO$_3$ superlattices, which demonstrate the formation of interfacial

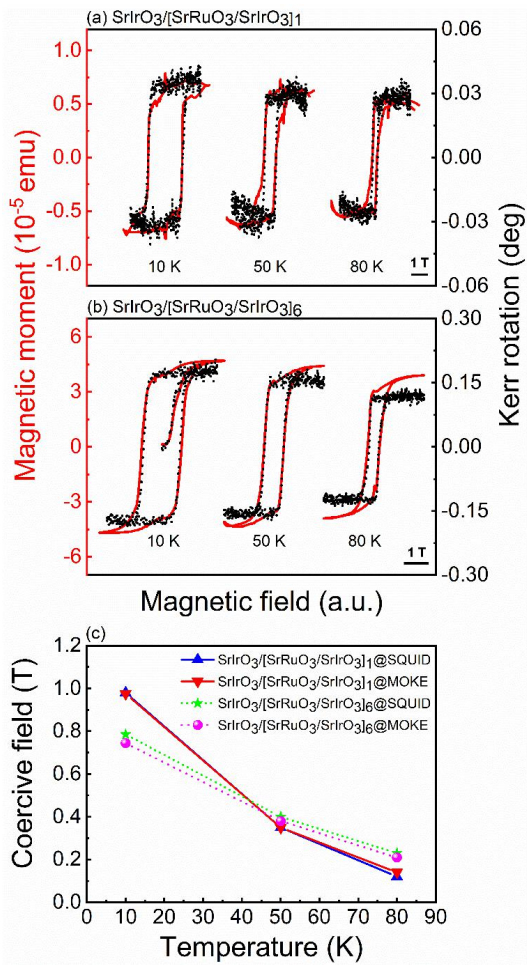

FIG. 3. Magnetic moment hysteresis loops (measured by SQUID magnetometry, red loops) and MOKE rotation angle loops for samples (a) $SrIrO_3/[SrRuO_3/SrIrO_3]_1$ and (b) $SrIrO_3/[SrRuO_3/SrIrO_3]_6$, measured in perpendicular magnetic field. (c) Comparison of the coercive fields of these two samples at different temperatures, as obtained from the SQUID and MOKE hysteresis loops. The lines are guide for the eye.

Ir-Mn molecular orbitals and ferromagnetic order of the Ir magnetic moments [43].

The comparison of the out-of-plane total magnetic moment hysteresis loops, measured with the SQUID magnetometer, and of the MOKE rotation angle loops of the samples with $m = 1$ and $m = 6$ is shown in Fig. 3(a) and Fig. 3(b), respectively. We plotted together the SQUID and MOKE hysteresis loops of the same sample at several temperatures (10 K, 50 K, 80 K). The coercive fields determined by SQUID and MOKE measurements are almost identical. The SQUID and MOKE loops of both samples are in very good agreement except for the regions of saturated magnetization, for relatively high fields. This discrepancy stems from the corrections that had to be applied to both type of loops: the loops are affected by different contributions either from the diamagnetic substrate and ferromagnetic impurities for SQUID loops or from the cryostat window in the case of the MOKE loops. The temperature dependence of the coercive field, extracted from loops measured by SQUID and MOKE, is compared for the two samples in Fig. 3(c). The magnitude of coercive fields and their temperature dependence is in good agreement with results of previous work [36].

### C. Anomalous Hall resistance and MOKE hysteresis loops

For the particular samples under study, the measured total Hall voltage has a contribution from the ordinary Hall effect and a contribution from the anomalous Hall effect, at temperatures below the Curie temperature of the $SrRuO_3$ layers. The total Hall voltage $V_{yx}$ was measured in van der Pauw configuration (as shown in the schematic inset of Fig. 4(a)). We define the transverse Hall resistance $R_{yx}$ as the ratio of the Hall voltage and the excitation current I: $R_{yx}= V_{yx}/I$.

For the $SrIrO_3/[SrRuO_3/SrIrO_3]_m$ multilayers, the metallic $SrRuO_3$ layers are magnetically decoupled [36] and are electrically connected in parallel. The $SrRuO_3$ layers have very similar resistances, because they have nominally the same thickness and similar interfaces [44]. The contribution of the ordinary Hall effect to the measured Hall voltage $V_{yx}$ was subtracted from all the Hall loops shown in the paper: we assumed that in the high magnetic field range, when the magnetization of the sample gets saturated, the only field dependence comes from the linear contribution of the ordinary Hall effect. Therefore, in the following $R_{yx}$ reflects the anomalous Hall effect of the samples and we refer to it as the anomalous Hall resistance in discussing the data presented in Fig. 4 and in the supplementary material.

The hysteresis loops of $R_{yx}$ at fixed temperature in the range 10 K-110 K/120 K of the $SrIrO_3/[SrRuO_3/SrIrO_3]_1$ and $SrIrO_3/[SrRuO_3/SrIrO_3]_6$ samples are plotted in Fig. 4(a) and Fig. 4(b), respectively. We note that the anomalous Hall resistance of these two samples exhibits a sign change from negative (at low temeperatures) to positive around 86 K. This change of sign is typical for $SrRuO_3$ epitaxial films as well as single crystals [45], consistent with previous experimental data and theoretical predictions [11, 46–51].This peculiar sign change, from negative to positive as the temperature increases, comes from the change of the sign of the intrinsic anomalous Hall conductivity. The latter is the result of the presence of Weyl like nodes, acting as magnetic monopoles, in the electronic band structure of $SrRuO_3$, combined with changes in the band structure as a function of the magnetization (and thus of the temperature). Although the existence of magnetic monopoles in $SrRuO_3$ is not experimentally unambiguously proved yet, the three-dimensional bulk $SrRuO_3$ has been considered as a system for which a large intrinsic AHE driven by topological band structure can be observed [45]. As the energies of nodal points and lines are different, when the Berry curvatures from them have opposite signs, the magnitude and the sign of intrinsic AHE conductivity can be tuned by changing the position of the Fermi level [49, 50, 52]. We note that measurements of a second trilayer sample, made in another PLD chamber, agree qualitatively with the AHE and MOKE loops data of the trilayer discussed here [see supplementary material, Fig. S3 and Fig. S5]. The most important observation is that the trilayer samples do not exhibit any humplike anomalies in the as measured Hall effect loops, as this is expected for symmetric interfaces.

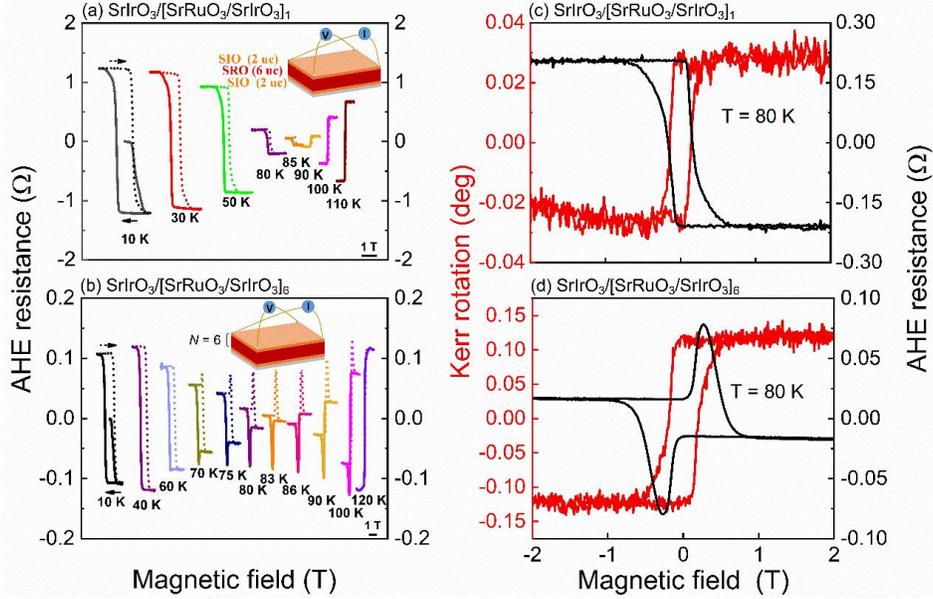

FIG. 4. Summary of the anomalous Hall effect (AHE) resistance $R_{yx}$ loops of the SrIrO$_3$/[SrRuO$_3$/SrIrO$_3$]$_m$ ($m$= 1, 6) samples, as a function of temperature: (a) for SrIrO$_3$/[SrRuO$_3$/SrIrO$_3$]$_1$ and (b) for SrIrO$_3$/[SrRuO$_3$/SrIrO$_3$]$_6$. In (c) and (d) the anomalous Hall resistance loops (black) and the MOKE rotation angle loops (red) measured at 80 K for sample SrIrO$_3$/[SrRuO$_3$/SrIrO$_3$]$_1$ and SrIrO$_3$/[SrRuO$_3$/SrIrO$_3$]$_6$, respectively, are compared.

Interestingly, the AHE resistance loops of the multilayer sample ($m$ = 6) do show humplike features within a broad temperature range from 70 K - 110 K (see **Fig.** 4(b)). These features of Hall resistance loops are peculiar, if compared with the corresponding magnetization/Kerr rotation angle loops measured at the same temperature. For the multilayer, the AHE and MOKE loops at the temperatures where the hump features occur are strikingly different, as obvious in the selected plots in **Fig.** 4(d). This indicates that the AHE resistance loops of the multilayer do not directly scale with the magnetization loops, as conventionally expected if the AHE constant was the same, in terms of magnitude and temperature dependence, for all the six SrRuO$_3$ layers of the multilayer.

We made a symmetric SrIrO$_3$/[SrRuO$_3$/SrIrO$_3$]$_m$ ($m$ = 3) multilayer in the second PLD system, with the same PLD parameters as for the second trilayer SrIrO$_3$/[SrRuO$_3$/SrIrO$_3$]$_1$. The AHE and MOKE loops of this multilayer with $m$ = 3, at different temperatures, are summarized in the supplementary material [see Fig. S4]. The behavior of the AHE loops was very intricate for this particular multilayer and humplike features occur in the temperature range of 10-80 K. We thus confirmed that the multilayers did have in common the appearance of the hump anomalies, in contrast to the bare trilayers. Our symmetric multilayers, SrIrO$_3$/[SrRuO$_3$/SrIrO$_3$]$_6$ and SrIrO$_3$/[SrRuO$_3$/SrIrO$_3$]$_3$, had a geometry that minimizes a net interfacial DMI. The lack of net DMI is a strong indication that other mechanisms than skyrmions and their topological Hall effect have to be considered for the hump anomalies of the AHE hysteresis loops. A more plausible explanation is that the individual SrRuO$_3$ layers of the multilayer SrIrO$_3$/[SrRuO$_3$/SrIrO$_3$]$_6$ have slightly different magnetic properties (i.e., saturation magnetization, coercive field, $T_c$), as a result of chemical and structural differences among each other (originating from slight layer thickness variation, different degree of intermixing of Ir on the Ru-site, and oxygen octahedron deformations). These differences, though probably minute, are of great importance for the temperature dependence and the magnitude of the intrinsic anomalous Hall resistivity of each layer. Hence, the individual ferromagnetic SrRuO$_3$ layers generate several independent magnetotransport channels leading to the observed hump-anomalies of the AHE loops. As proposed in several papers [28, 33, 52–54] and in our previous reports [32, 35, 38], the humplike anomalies of the AHE hysteresis loops in SrRuO$_3$-based heterostructure can be well explained by a model of several independent magnetic channels, with distinct coercive fields and different temperatures at which the intrinsic AHE conductivity changes sign.

## IV. SUMMARY

In epitaxial asymmetric $SrRuO_3/SrIrO_3$ bilayers a strong interfacial Dzyaloshinskii-Moriya interaction (DMI) was proposed to emerge and to be the driving force for the formation of skyrmions. These skyrmions would result in a topological Hall effect, whose manifestation was considered to be spotted as humplike features, developing while the magnetization of the $SrRuO_3$ layer reversed between saturated states. We studied here heterostructures in which an ultrathin ferromagnetic $SrRuO_3$ layer was sandwiched between $SrIrO_3$ layers. Principally, this geometry disfavors the occurrence of a net interfacial DMI and thus the formation of skyrmions would be exceptional. $SrIrO_3/SrRuO_3/SrIrO_3$ trilayers did not have hump anomalies of the Hall resistance loops. However, the Hall resistance loops of multilayers, in which the trilayer was stacked several times, did exhibit the humplike structures, similar to the asymmetric $SrRuO_3/SrIrO_3$ bilayers. The magnetization as a function of temperature indicated that the multilayers had a spread of the Curie temperatures, hinting to differences in the magnetic properties of the individual $SrRuO_3$ layers. The origin of the Hall effect anomalies likely stems from unavoidable structural differences between the individual $SrRuO_3$ layers stacked in epitaxial multilayers. The minute structural differences (oxygen octahedra rotation angles, bond lengths) of the individual ruthenate layers result in inhomogeneous magnetic and electrical properties across the multilayer. It is possible that the individual $SrRuO_3$ layers generate several independent magnetotransport channels leading to the observed anomalous features of the Hall effect loops. The relation of the hump anomalies to the skyrmion formation cannot be ruled out, however our data strongly support the interpretation in terms of multiple magnetotransport channels present in multilayers.

## V. ACKNOWLEDGEMENT

We thank Michael Ziese for constant valuable advice with the physical properties of $SrRuO_3$ samples and with the SQUID and Hall measurements (University of Leipzig). We are grateful to Achim Rosch for fruitful discussions and insightful suggestions, Susanne Heijligen for kind assistance with SQUID measurements, and Andrea Bliesener for AFM and assistance with the PPMS measurements (University of Cologne). We thank René Borowski for etching the STO substrates (FZ Julich). This work was supported by the German Research foundation (DFG) (projects number 335038432 and 403504808) and through CRC1238 (Project No. 277146847). PvL and ILV thank DFG for funding the purchase of the PLD-RHEED system at University of Cologne, with which some of the investigated samples were grown (Project No.407456390). Also support from the German Excellence Initiative via the key profile area "quantum matter and materials" (QM2) of the University of Cologne is gratefully acknowledged. L. Y. thanks for financial support from the China Scholarship Council. (File No.201706750015).

*Correspondence to*: L.Y. (yanglin@ph2.uni-koeln.de) and I. L.-V. (vrejoiu@ph2.uni-koeln.de)

# Supplementary online materials
# Origin of the hump anomalies in the Hall resistance loops of ultrathin SrRuO$_3$/SrIrO$_3$ multilayers

Lin Yang,[1] Lena Wysocki,[1] Jörg Schöpf,[1] Lei Jin,[2] András Kovács,[2] Felix Gunkel,[3] Regina Dittmann,[3] Paul H. M. van Loosdrecht,[1] and Ionela Lindfors-Vrejoiu[1]

[1]*University of Cologne, Institute of Physics II, 50937 Cologne, Germany*
[2]*Ernst Ruska-Centre for Microscopy and Spectroscopy with Electrons, Forschungszentrum Jülich GmbH, 52425 Jülich, Germany*
[3]*PGI-7, Forschungszentrum Jülich GmbH, 52428 Jülich, Germany*

Dated: September 22, 2020

## I. STRUCTURAL CHARACTERIZATION: IN SITU RHEED AND AFM INVESTIGATIONS

We monitored the growth mode and the thickness of the individual layers of the multilayers by employing *in situ* high oxygen pressure reflective high-energy diffraction (RHEED). The average intensity of the RHEED specular spot as a function of time is shown in **Fig. S1**(a) and **Fig. S1**(c) for SrIrO$_3$/[SrRuO$_3$/SrIrO$_3$]$_1$ and SrIrO$_3$/[SrRuO$_3$/SrIrO$_3$]$_6$, respectively, both grown in the same PLD chamber at FZ Jülich under the same PLD parameters. The oscillations in the RHEED intensity-time curves during the SrIrO$_3$ deposition show that the iridate layers grew in a layer-by-layer growth mode: see a first clear oscillation followed by a more damped second oscillation, marked by the small black arrows, corresponding to the growth of two monolayers of SrIrO$_3$. SrRuO$_3$ layers grew in step-flow growth regime [1]. The similar RHEED intensity behavior indicates the homogeneous thickness of sample layers. The surface morphology of both SrIrO$_3$/[SrRuO$_3$/SrIrO$_3$]$_1$ and SrIrO$_3$/[SrRuO$_3$/SrIrO$_3$]$_6$ samples was investigated by atomic force microscopy (AFM) and is shown in **Fig. S1**(b) and **Fig. S1**(d), respectively. Non-continuous terrace-like structure is shown by the trilayer (see **Fig. S1**(b)) and the multilayer has a rather large density of tiny holes, coming from probably incomplete coverage of the top most SrIrO$_3$ layer (see **Fig. S1**(d)).

As shown in the main paper a SrIrO$_3$/[SrRuO$_3$/SrIrO$_3$]$_6$ multilayer was investigated by scanning transmission electron microscopy (STEM) in the high-angle annular dark field mode (HAADF) (see Fig. 1 of the paper). We made energy dispersive x-ray spectroscopy (EDX) maps of the chemical elements Sr, Ti, Ru and Ir across the multilayer, as shown in **Fig. S2**(a). Line profiles for each element were acquired and are shown in **Fig. S2**(b): they confirm that the SrRuO$_3$ layers are about 6-7 uc thick (as the number of individual Ru-O$_2$ planes varies between 6 and 7), while the SrIrO$_3$ layers are 2-3 uc thick (as the number of individual Ir-O$_2$ planes varies between 2 and 3), in agreement with our RHEED observations. In **Fig. S2**(b), we drew four lines, two in orange and the other two in green. The first orange (green) line marks the position of Ir (Ru), the second orange (green) line marks the neighboring B-site positions. For both cases, the net count drops from 80 to 30. That means no matter what the element is, the decrease is the same. Only the thickness of the individual layers in the growth direction matters. The SrIrO$_3$ layers, being only 2-3 uc thick, are much too thin to allow quantitative analyses of the possible intermixing at the interfaces with the SrRuO$_3$ layers. To achieve atomic resolution in EDX investigations, due to the electron beam channeling, volume and spectrum background effects, is very problematic.

## II. MOKE AND HALL LOOPS OF A SECOND SrIrO$_3$/[SrRuO$_3$/SrIrO$_3$]$_1$ TRILAYER AND A SrIrO$_3$/[SrRuO$_3$/SrIrO$_3$]$_3$ MULTILAYER MADE IN A SECOND PLD SYSTEM

To investigate the reproducibility of the behavior of the AHE loops, we grew the trilayer sample SrIrO$_3$/[SrRuO$_3$/SrIrO$_3$]$_1$ and a multilayer sample with $m = 3$ SrIrO$_3$/[SrRuO$_3$/SrIrO$_3$]$_3$, in our PLD system at University of Cologne. The MOKE and Hall effect loops of the samples are summarized in **Fig. S3** and in **Fig. S4**, respectively. The Kerr rotation angle and AHE resistance measurements were performed simultaneously and both type of loops show similar coercive fields at all temperatures. In the ferromagnetic phase of SrRuO$_3$ layer, as the temperature increases, the AHE resistance changes sign, from negative to positive, somewhat below 80 K. There are no humplike features in the AHE resistance loops also for this second SrIrO$_3$/[SrRuO$_3$/SrIrO$_3$]$_1$. These results are consistent with the trilayer sample reported in the main paper (see Fig. 4 of the paper).

For the multilayer SrIrO$_3$/[SrRuO$_3$/SrIrO$_3$]$_3$ (see **Fig. S4**), the MOKE and Hall loops show striking differences in the temperature range 10 K - 80 K. For the MOKE measurement, the open MOKE loops are obtained up to about 110 K, indicating that the Curie temperature is at least 110 K. For the anomalous Hall effect resistance measurement, the clear humps exist from the lowest temperature we can measure at, 10 K, to about 80 K. The evolution of the sign of anomalous Hall constant is quite different from the previous sample SrIrO$_3$/[SrRuO$_3$/SrIrO$_3$]$_6$ and bilayer sample in Matsuno *et al.* paper [2]. The sign of the total anomalous Hall resistance (voltage) of the sample SrIrO$_3$/[SrRuO$_3$/SrIrO$_3$]$_3$ is positive down to about 10 K. The multiple peaks of the AHE loops at 10 K and 30 K may be explained, if the global loop is decomposed in three independent loops generated by the three SrRuO$_3$ layers. The three separated magnetic layers possess slightly different magnetic and AHE properties [3].

### III. AHE RESISTANCE AND MOKE ROTATION ANGLE FIELD LOOPS FOR TWO SYMMETRIC SrIrO$_3$/[SrRuO$_3$/SrIrO$_3$]$_1$ TRILAYERS

The data of AHE resistance and MOKE loop measurement at different temperatures for the trilayer made in another PLD system and its reference trilayer sample (studied in the main paper) are summarized in **Fig.** S5. It should be noted that we obtained these data by simultaneous measurement of MOKE and Hall effect resistance loops in our combined MOKE-Hall setup, with the sample in the same cryostat. In general, for each sample, the AHE loops scale with the MOKE loops fairly well. The most striking differences between the two samples are the magnitude of the coercive fields and the temperature dependence of the anomalous Hall effect resistance. The second trilayer has much smaller coercive field at low temperatures, see for instance the loops measured at 10 K: the coercive field of the reference layer is almost twice as large. The AHE changes sign well below 80 K for the second trilayer, while the reference trilayer changes the sign of the AHE from negative to positive at 86 K. We grew the second trilayer with the intention to obtain a fairly comparable sample, i.e. with 2 uc thick SrIrO$_3$ and 6 uc thick SrRuO$_3$ layers. However, the second PLD chamber has major differences (such as target-to-substrate distance, perpendicular geometry, laser fluence measurement), which made it not possible to have the same PLD parameters for the growth. Thus, we stress how important the growth conditions are for the magnetic and electronic properties of SrRuO$_3$/SrIrO$_3$ oxide thin films. However, the most important similarity between the two trilayer samples is that both do not show any humplike anomalies of the AHE resistance loops, demonstrating a consistent behavior for the expected symmetric trilayers.

*Correspondence to*: L.Y. (yanglin@ph2.uni-koeln.de) and I. L.-V. (vrejoiu@ph2.uni-koeln.de)

---

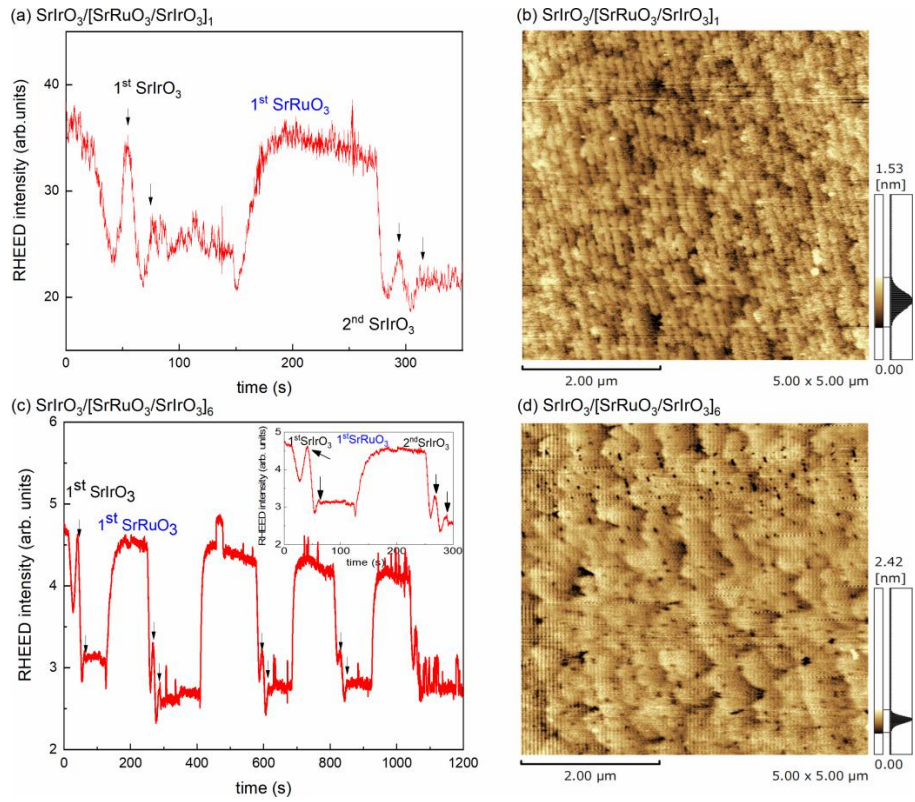

FIG. S1. RHEED and AFM investigations of $SrIrO_3/[SrRuO_3/SrIrO_3]_1$ and $SrIrO_3/[SrRuO_3/SrIrO_3]_6$, which were grown under the same PLD conditions: (a) and (c) deposition time dependence of RHEED intensity; (b) and (d) AFM topography images (5 $\mu$m × 5 $\mu$m scans) of the top surface of the as grown samples. The small black arrows in (a) and (c) mark the top of the oscillations of the RHEED intensity signal during the growth of the $SrIrO_3$ layers. The inset in (c) shows the RHEED signal recorded during the growth of the first three layers of the multilayer $SrIrO_3/[SrRuO_3/SrIrO_3]_6$.

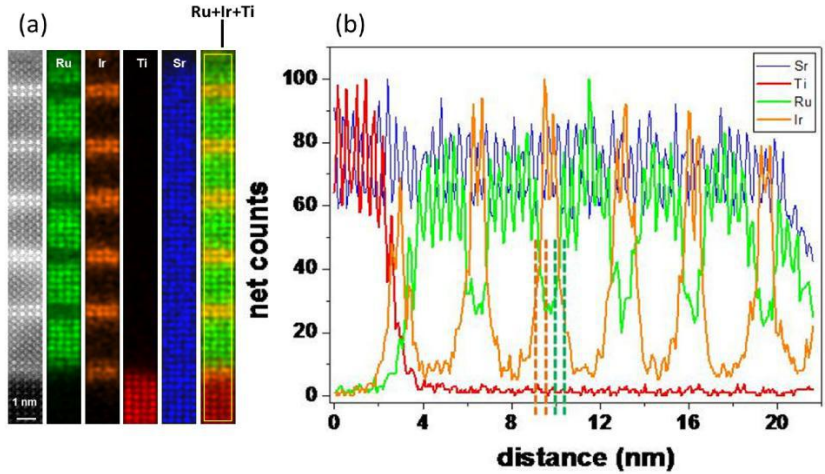

FIG. S2. (a) HAADF-STEM and EDX analyses of a $SrIrO_3/[SrRuO_3/SrIrO_3]_6$ multilayer. The line profiles across the multilayer, starting from the substrate upwards in the growth direction, for the atomic column with the Sr, Ti, Ru, Ir ions are shown in (b).

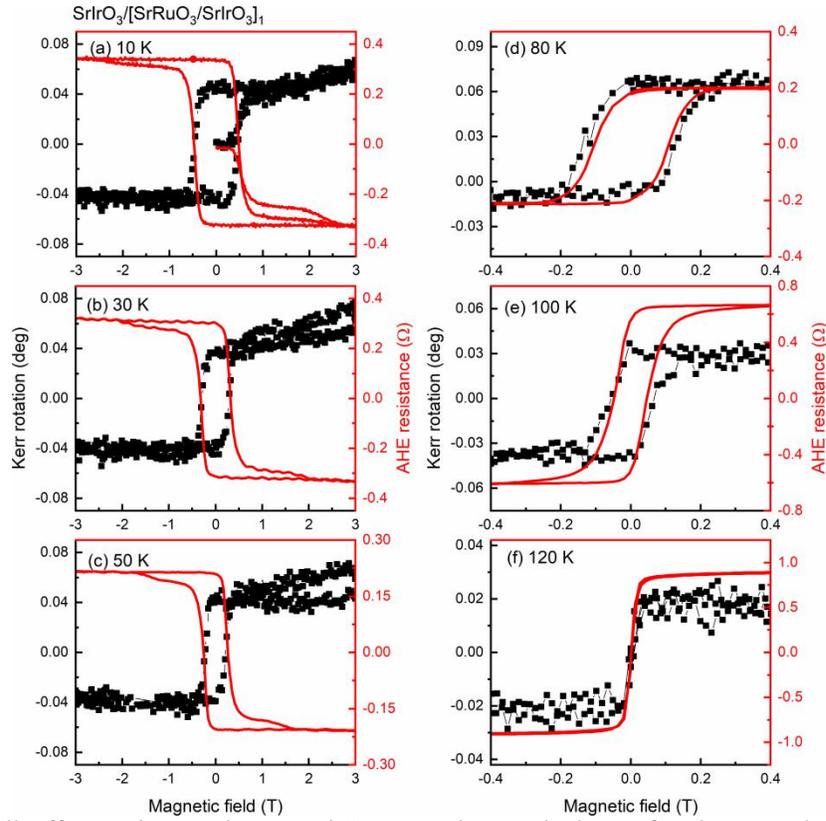

FIG. S3. Anomalous Hall effect resistance loops and Kerr rotation angle loops for the second SrIrO$_3$/[SrRuO$_3$/SrIrO$_3$]$_1$ trilayer at different temperatures from 10 K to 120 K: AHE resistance loops (red line) and the MOKE loops (black line with solid square dots). The AHE changes sign to positive above 60-70 K.

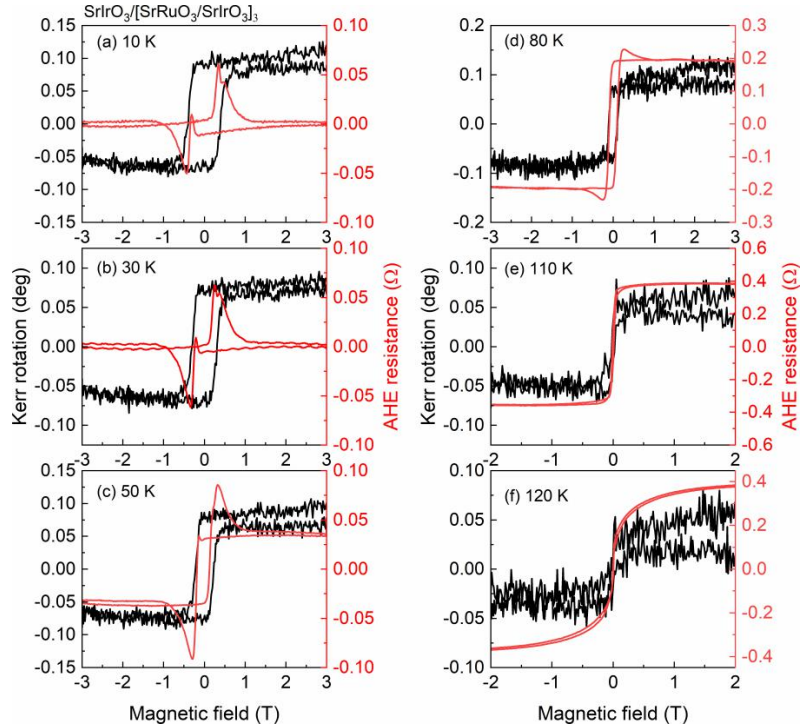

FIG. S4. Anomalous Hall effect resistance loops and Kerr rotation angle loops for the multilayer SrIrO$_3$/[SrRuO$_3$/SrIrO$_3$]$_3$ ($m = 3$), at different temperatures from 10 K to 120 K: AHE resistance loops (red line) and the MOKE loops (black line).

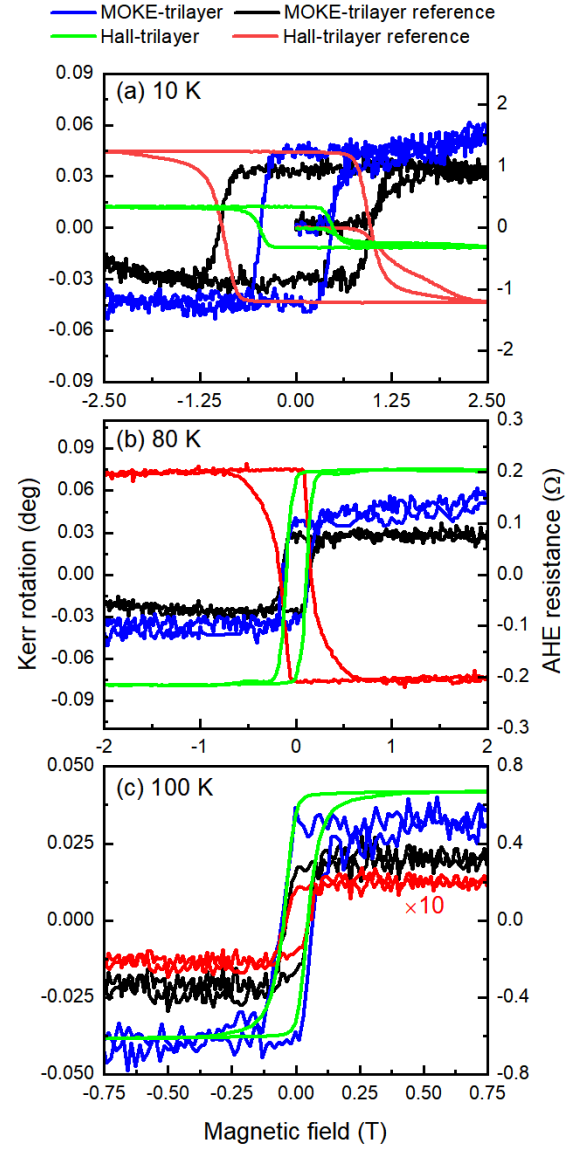

FIG. S5. Comparison of AHE and MOKE loops for the two $SrIrO_3/[SrRuO_3/SrIrO_3]_1$ trilayers (the *reference* is the trilayer presented in the main paper), made in different PLD chambers, with differing PLD conditions. We compare the loops at different temperatures, capturing the change of sign of AHE for both samples: (a) 10 K, (b) 80 K, (c) 100 K. Anomalous Hall effect (AHE) resistance loops are plotted in red and green and the Kerr rotation angle black and blue. The magnitude of Hall loop for $SrIrO_3/[SrRuO_3/SrIrO_3]_1$ at 100 K was increased tenfold for better comparison.